\documentclass[10pt]{amsart}
     \makeatletter
     \def\section{\@startsection{section}{1}%
     \z@{.7\linespacing\@plus\linespacing}{.5\linespacing}%
     {\bfseries%\normalfont\scshape
     \centering
     }}
     \def\@secnumfont{\bfseries}
     \makeatother
\newtheorem{theorem}{Theorem}[section]
\newtheorem{lemma}[theorem]{Lemma}

\newtheorem{corollary}[theorem]{Corollary}
\theoremstyle{definition}

\newtheorem{example}[theorem]{Example}
\newtheorem{assumption}[theorem]{Assumption}
\theoremstyle{remark}

\numberwithin{equation}{section}
\usepackage{amsmath,amsthm,amssymb,amsbsy}

\usepackage[all]{xy}
\usepackage{chngcntr}
\usepackage{url}
\usepackage{graphicx}

\usepackage{tikz}
\usetikzlibrary{arrows,positioning,calc,shapes.multipart}

\tikzset{
    >=stealth',
    markovnode/.style ={rectangle, rounded corners, draw=black, very thick, text width=6.5em, minimum height=3em, text centered},
    markovAuxNode/.style ={rectangle, rounded corners, draw=white, very thick, text width=6.5em, minimum height=3em, text centered},
    markovnodeSmall/.style ={rectangle, rounded corners, draw=black, very thick, text width=3.5em, minimum height=2em, text centered},
    markovAuxNodeSmall/.style ={rectangle, rounded corners, draw=white, very thick, text width=3.5em, minimum height=2em, text centered},
    markovarrow/.style={->, thick, shorten <=2pt, shorten >=2pt},
    markovarrowdashed/.style={->, dashed, shorten <=2pt, shorten >=2pt}
    }

\counterwithin{figure}{section}

\newcommand{\RR}{\mathbb{R}}

\newcommand{\BBB}{\mathcal{B}}

\newcommand{\FFF}{\mathcal{F}}

\newcommand{\dv}{\,\mathrm{d}}                         % Integrator
\newcommand{\fvsp}{\textbf{\upshape FV}}                      % fv for finite variation

\begin{document}

\setlength{\parindent}{0cm}
\setlength{\parskip}{0.5cm}

\title[A generic model for spouse's pensions]{A generic model for spouse's pensions with a view towards the calculation of liabilities}

\date{\today}

\author[A. Sokol]{Alexander Sokol}

\address{Alexander Sokol: Institute of Mathematics, University of
  Copenhagen, 2100 Copenhagen, Denmark, alexander@math.ku.dk}

\subjclass[2010] {Primary 62P05; Secondary 60G55}

\keywords{Life insurance, Liability, Cashflow, Marked point process, Longevity}

\begin{abstract}
We introduce a generic model for spouse's pensions. The generic model allows for
the modeling of various types of spouse's pensions with payments commencing at the
death of the insured. We derive abstract formulas for cashflows and liabilities
corresponding to common types of spouse's pensions. We show how the standard
formulas from the Danish G82 concession can be obtained as a special case of our generic
model. We also derive expressions for liabilities for spouse's pensions in models more advanced
than found in the G82 concession. The generic nature of our model and results furthermore
enable the calculation of cashflows and liabilities using simple estimates of marital
behaviour among a population.
\end{abstract}

\maketitle

\noindent

\section{Introduction}

\label{sec:SpouseIntro}

The motivation for this paper is the accurate calculation of the liabilities corresponding
to the particular type of life insurance policies known as spouse's pensions. In such a policy,
payments are made to the spouse upon the death of the insured, in the case where a spouse is
present. Many pension funds offer products such as this and have a considerable interest in
efficient practical estimation of their corresponding liabilities to the policyholders.

In the Danish G82 concession, specifying many standard types of life insurance policies, several types of
spouse's pensions are described, see e.g. \cite{CederbyeEtAl1997} for more on this. The G82 concession
describes specific formulas for the calculation of the liability, meaning the expectation
of the discounted value of the future payments, under certain assumptions. The
formulas, however, are built around a very particular and unwieldy model, for which recent estimates
of the parameters are not generally known. Our main objective in this paper is to develop a flexible
modeling framework for estimation of liabilities for spouse's pensions.

Apart from this, the forthcoming Solvency II rules from the European Union has led to increased
theoretical and practical interest in the calculation of not only the liability, but also the cashflow
of insurance policies, meaning the expected rate of payments on the insurance policy in the future, see e.g. \cite{BuchardMoellerSchmidt2013}. One classical setup for such
calculations is to let e.g. the health state of the insured be modeled by a continuous-time Markov
chain or semi-Markov chain, see \cite{Hoem1969,Hoem1972}. For spouse's pensions, the presence of a future spouse with a priori
unknown age excludes the possibility of a simple Markov chain model, and therefore different methods
must be applied to obtain expressions for the cashflow of such policies. Consistently with the above,
we aim to obtain a modeling framework which enables the calculation of both liabilities and cashflows.

%ChristiansenMultistate2012,

Finally, the Solvency II rules also specify the necessity of modeling the longevity risk inherent
in life insurance products, meaning the modeling of longevity improvements in populations over time. In
the liability expressions of the G82 concession, longevity improvements are not present. It is therefore
of interest to obtain models for the calculation of the liability of spouse's pensions in which
longevity improvements are included, such that e.g. the mortality benchmark intensities with
longevity improvements reported in \cite{FSALongevity} can be used when calculating cashflows and liabilities.

In this article, we develop a generic model for spouse's pensions, and derive expressions for
cashflows and liabilities for a wide family of pension products. We also show how to obtain the
classical G82 concession expressions for the liability corresponding to a spouse's pension as
a special case, using a marked point process model. Finally, we show how to extend this model to include
longevity improvements.

The remainder of the article is structured as follows. In Section \ref{sec:MCReview}, we review
the notions of payment processes, expected cumulative payments, cashflows and liabilities in the context of a simple Markov chain model.
In Section \ref{sec:generic}, we introduce our generic model for spouse's pensions and derive expressions for cashflows and liabilities. In
Section \ref{sec:LongevityMPP}, we show how the expressions from the G82 concession can be replicated
in our framework through a marked point process model. Here, we also show how to extend this model
to include longevity effects. Finally, in Section \ref{sec:Discussion}, we discuss our results.
Appendix \ref{sec:Proofs} contains proofs.

\section{Review of the continous-time Markov chain framework}

\label{sec:MCReview}

In order to motivate our model, we first recall the modeling framework based on continuous-time
Markov chains as discussed in e.g. \cite{BuchardMoellerSchmidt2013}. Consider a simple life insurance product
paying one amount of monies per time from a given timepoint $c$ and onwards, for as long as the
insured is alive. Let $Z$ be the health state of the insured, taking the values $a$ (alive)
and $d$ (dead). In order to model this insurance product, we may consider the process
\begin{align}
  B_t &= \int_0^t 1_{(Z_t = a)}1_{(t \ge c)}\dv s.
\end{align}
For any $t\ge0$, $B_t$ describes the cumulative payments paid out to the insured. We refer to $B$
as the cumulative payments process, or simply as the payment process. We may then consider
\begin{align}
\label{eq:MCExpectedCumPay}
  A_t &= EB_t = \int_0^t 1_{(t\ge c)}P(Z_t = a)\dv s,
\end{align}
the expected cumulative payments. Since $A$ is continous and differentiable almost everywhere,
we may let $a$ denote the Radon-Nikodym derivative with respect to the Lebesgue measure, yielding
\begin{align}
\label{eq:MCCashflow}
  a_t &= 1_{(t\ge c)}P(Z_t = a).
\end{align}
We refer to $a$ as the cashflow corresponding to the
insurance policy. Finally, introducing an interest rate model based on a deterministic
short rate $r$, we may define
\begin{align}
\label{eq:MCLiability}
  L &= E \int_0^\infty e^{-rt} \dv B_t,
\end{align}
the liability corresponding to the insurance policy. These concepts of cumulative payment processes,
expected cumulative payments, cashflows and liabilities, are well known in various guises from the literature,
see e.g. \cite{Hoem1969,Hoem1972,Norberg1991,BuchardMoellerSchmidt2013}. In the next section, we use the same framework in the context of a generic model for spouse's pensions.

\section{A generic model for spouse's pensions}

\label{sec:generic}

In this section, we introduce our generic model for spouse's pensions. We are interested
in modeling spouse's pensions of the type where the spouse is entitled to certain payments
contingent upon the death of the insured as well contingent upon a generic "policy state".
Usually, this latter "policy state" will be the health state of the spouse, such that e.g.
payments only are made for as long as the spouse is alive, but for flexibility, we do not
limit ourselves as regards the nature of this policy state space. After the introduction of
the modeling framework, we derive expressions for cashflows and liabilities in the generic model.
Also, we illustrate the usefulness of our model by deriving expressions for cashflows
and liabilities for several types of spouse's pensions.

Assume given a probability space $(\Omega,\FFF,P)$. Let $T$ be a random variable taking its values in $\RR_+$, describing
the time of death for the insured. Let $X$ be a random variable
taking the two values $x_s$ and $x_m$, corresponding to "single" and "married",
respectively, describing the marital state of the insured at the time of death, $T$.
Let $Y$ be a random variable denoting the age of the spouse at the time of death $T$. Here, we let $\partial\notin\RR_+$
be a "coffin state" held by $Y$ if the insured was unmarried at the time of death,
meaning that we assume $Y=\partial$ whenever $X = x_s$, and otherwise $Y$ takes its values in $\RR_+$.

Finally, for each $u,y\ge0$, let $(Z^{u,y}_r)_{r\ge u-y}$ denote a stochastic process on $[u-y,\infty)$ with some common finite
state space $E$. We think of $Z^{u,y}$ as a stochastic generic "policy state" for
the case where a spouse exists at the time of death of the insured, and that spouse has age $y$
at time $u$. Consistently with this, we let $Z^{u,y}$ be defined on $[u-y,\infty)$, where
$u-y$ is the (possibly negative) timepoint when the spouse then had age zero. In the most common case,
$Z^{u,y}$ will describe the health of the spouse. Furthermore, let $D_r(E)$ denote the space
of cadlag functions from $[r,\infty)$ to $E$, and assume that $Z^{u,y}$ takes its values in $D_{u-y}(E)$. Let $\fvsp$
denote the space of mappings of finite variation from $\RR_+$ to $\RR$. For $u,y\in\RR_+$,
assume given measurable mappings $\Pi_{u,y}:D_{u-y}(E)\to\fvsp$, where both spaces are endowed with
the $\sigma$-algebras induced by the coordinate projections. We refer to the mappings $(\Pi_{u,y})$
as the payment functions, and define a process $C$ by
\begin{align}
  C_t &= \Pi_{T,Y,t}(Z^{T,Y}),
\end{align}
where $\Pi_{u,y,t}(z) = \Pi_{u,y}(z)_t$. The interpretation of this is as follows. The expression $\Pi_{u,y,t}(z)$ represents
the cumulative payments made to the spouse at time $t$, given that the death
of the insured occurred at time $u$, that the insured was married at that time, and that the
age of the spouse at that time was $y$, and given the policy state history $z$ since the
birth of the spouse. As a consequence, $C$ represents the unconditional cumulative payments for the insurance policy
with payment functions $(\Pi_{u,y})$, excepting that $C$ does not prescribe payments to begin
conditionally upon the death of the insured while married.

It remains to define the actual cumulative payment process corresponding to the components of the spouse
pension described above, similarly to how we in Section \ref{sec:MCReview} defined cumulative payment processes
for a simple policy. To this end, we define a process $B$ by
\begin{align}
\label{eq:PensionCPP}
 B_t &= \int_0^t 1_{(s\ge T)} 1_{(X = x_m)} \dv C_s.
\end{align}
The process $B$ then has paths of finite variation, and corresponds to the cumulative payment process
for the spouse's pension with payment functions $(\Pi_{u,y})$. Given the joint distribution of
$(T,X,Y,Z^{T,Y})$, the model for spouse's pensions is fully specified, and given the payment functions $(\Pi_{u,y})$,
a particular spouse's insurance policy is fully specified.

\begin{example}
\label{example:G810}
We will show how the above framework contains a classical spouse's pension, corresponding to
the G82 concession reward no. 810. Let $E=\{a,d\}$, where $a$ denotes alive and $d$ denotes dead,
corresponding to that $Z^{u,y}_r$ denotes the health state of the spouse at time $r\ge u-y$, for the
case where the spouse has age $y$ at time $u$. Define, for $t\ge 0$,
\begin{align}
\label{example:G810Payment}
  \Pi_{y,u,t}(z) &= 1_{(y\neq\partial)}1_{(t\ge u)}\int_u^t 1_{(z_r=a)}\dv r.
\end{align}
This choice of $\Pi$ corresponds to the spouse receiving monies at a rate of one unit of money per
unit of time, whenever the spouse is alive, starting at the death of the insured at time $u$. To see this, note that
\begin{align}
  C_t &= \Pi_{T,Y,t}(Z^{T,Y})
       = 1_{(Y\neq\partial)}1_{(t\ge T)} \int_T^t 1_{(Z^{T,Y}_r=a)}\dv r,
\end{align}
corresponding to $C_t$ being equal to monies accumulated at a rate of one unit per unit of time,
over the time period $[T,t]$ from the time of death of the insured to the current time $t$,
with payments only accumulating at time $r\in[T,t]$ when the spouse is alive. The presence
of $Z^{T,Y}$ in the integral corresponds to the age of the spouse at time $T$ being $Y$. Note
that we need to include the indicator for $y\neq\partial$ in the definition in order to ensure
that the expression for $\Pi$ is well-defined even when $y=\partial$. In practice, we only
take interest in the values of $C_t$ when $Y\neq\partial$, so the actual values of $\Pi$
when $y=\partial$ is not of any consequence. We can see this by considering the cumulative payment
process of the insurance policy, which is
\begin{align}
\label{eq:G810CPP}
 B_t &= \int_0^t 1_{(s\ge T)} 1_{(X = x_m)} \dv C_s
      = 1_{(t\ge T)} 1_{(X = x_m)} \int_T^t 1_{(Z^{T,Y}_r=a)}\dv r,
\end{align}
since $Y\neq \partial$ whenver $X=x_m$.
\hfill$\circ$
\end{example}

As in Section \ref{sec:MCReview}, we may now take an interest in the cashflow and liability corresponding
to the cumulative payment process (\ref{eq:PensionCPP}). In order to prove expressions for these,
we require some regularity conditions on the joint distribution of the variables $T$, $X$ and $Y$ and the
processes $Z^{u,y}$, as well as regularity conditions on the payment mappings.

\begin{assumption}
\label{ass:Generic}
We assume that the distribution of $T$ has a density $h$ with respect to the Lebesgue measure on $\RR_+$, that the
conditional distribution of $Y$ given $T=u$ and $X=x_m$ has a density $f(\cdot|u)$ with respect
to the Lebesgue measure on $\RR_+$, and that each policy state process $Z^{u,y}$ is independent of $(T,X,Y)$. Finally,
we assume that for $u>0$, $y\ge0$ and $z\in D_{u-y}(E)$, it holds that $\Pi_{u,y,u-}(z)=0$.
\end{assumption}

Here, $\Pi_{u,y,u-}(z)$ is the limit of $\Pi_{u,y,r}(z)$ as $r$ tends to $u$ from below. The latter
assumption that $\Pi_{u,y,u-}(z)=0$ corresponds to assuming that no payments are made prior
to the death of the insured. Removing this assumption would not complicate matters considerably, but would
merely result in different factors in e.g. the formula (\ref{eq:CumulativeCashflow}). However, as the
condition $\Pi_{u,y,u-}(z)=0$ holds for all spouse's pensions of interest to us, we make the assumption in
order to simplify our results.

In the remainder of this section, we assume that Assumption \ref{ass:Generic} is in force. Furthermore,
we let $g(u)=P(X=x_m|T=u)$ for $u\ge0$. The following three results yield expressions for the expected
cumulative payments, the cashflow and the liability of a spouse's pension, respectively, corresponding
to (\ref{eq:MCExpectedCumPay}), (\ref{eq:MCCashflow}) and (\ref{eq:MCLiability}) of the example in
Section \ref{sec:MCReview}.

\begin{theorem}
\label{theorem:ECCFormula}
It holds that
\begin{align}
\label{eq:CumulativeCashflow}
  A_t  &=\int_0^t h(u)g(u) \int_0^\infty f(y|u) E\Pi_{u,y,t}(Z^{u,y}) \dv y\dv u.
\end{align}
\end{theorem}

The fomula (\ref{eq:CumulativeCashflow}) may be interpreted as follows. When calculating the expected
cumulative payments of the spouse's pension, we first condition on the time of death while married, which
has density $u\mapsto h(u)g(u)$. Note that as this event does not always occur, the density
$u\mapsto h(u)g(u)$ is generally defective, meaning that it does not have unit Lebesgue integral, instead its
integral over $\RR_+$ is equal to $P(X = x_m)$. Having conditioned on the time of the death of the insured while married, (\ref{eq:CumulativeCashflow})
yields that the expected value of the cumulative payments up to time $t$ is
\begin{align}
\label{eq:CCGivenDeathWhileMarried}
  \int_0^\infty f(y|u) E\Pi_{y,u,t}(Z^{u,y}) \dv y\dv u.
\end{align}
This corresponds to first conditioning upon the age of the spouse at the time of death of the insured, which
has density $f(\cdot|u)$ when the time of death is $u$. Given this age, the formula (\ref{eq:CCGivenDeathWhileMarried})
then shows that the expected value of the cumulative payments up to time $t$ is
\begin{align}
\label{eq:CCGivenDeathWhileMarriedAndSpouseAge}
   E\Pi_{y,u,t}(Z^{u,y})
\end{align}
which corresponds to expected value of the payments $\Pi_{y,u,t}(Z^{u,y})$ accumulated from time $u$, when the death of
the insured occurred, to time $t$. Here, the independence of $Z^{u,y}$ and $(T,X,Y)$ ensures that the expectation does not
need to be conditioned upon the values of $T$, $X$ and $Y$.

\begin{corollary}
\label{corr:SpousePensionCashflow}
Assume that $\Pi_{y,u,u}(z) = 0$ and that for all $y,u\ge0$, the mapping $t\mapsto E\Pi_{y,u,t}(Z)$ is differentiable
for $t>u$ with a derivative which is bounded on compact sets. It then holds that the cashflow $(a_t)$
corresponding to the cumulative payment process exists, and it is given by
\begin{align}
\label{eq:Cashflow}
 a_t  &= \int_0^t h(u)g(u) \int_0^\infty f(y|u) \left(\frac{\dv}{\dv t} E \Pi_{u,y,t}(Z^{u,y})\right) \dv y\dv u.
\end{align}
\end{corollary}

\begin{corollary}
\label{corr:SpousePensionLiability}
Assume the regularity conditions of Corollary \ref{corr:SpousePensionCashflow}. Given a deterministic short rate $r$, the liability corresponding to the payment functions
$(\Pi_{y,u,t})$ is given by
\begin{align}
\label{eq:Liability}
 L  &= \int_0^\infty h(u)g(u) \int_0^\infty f(y|u) \int_u^\infty  e^{-rt} \left(\frac{\dv}{\dv t} E \Pi_{u,y,t}(Z^{u,y})\right) \dv t \dv y \dv u.
\end{align}
\end{corollary}

In the corollaries, (\ref{eq:Cashflow}) is essentially the same as (\ref{eq:CumulativeCashflow}) after differentiation
under the integral sign, resulting in the dependency of the cashflow upon the conditional payments $E\Pi_{y,u,t}(Z)$
is solely through the rate of payment, meaning the derivative of $t\mapsto E\Pi_{y,u,t}(Z)$. As for (\ref{eq:Liability}),
this is simply the discounted expected payments after rearrangement of the order of integration.

For most policies of interest, the regularity conditions of Corollary \ref{corr:SpousePensionCashflow} and
Corollary \ref{corr:SpousePensionLiability} will be satisfied. Before proceeding to the next section, we give some examples
of how to apply the corollaries in order to obtain expressions for cashflows and liabilities for concrete types of
spouse's pensions. In Example \ref{example:G810Liability}, we derive a G82-type formula for a classical spouse's
pension, corresponding to the G82 reward no. 810. In Example \ref{example:G815Longevity}, we calculate the liability
of a terminating spouse's pension, corresponding to the G82 reward no. 810, in a model where the health state $Z^{u,y}$
of the spouse is modeled with longevity improvements. Finally, in Example \ref{example:LumpSumLongevity}, we consider
a reward outside the scope of the G82 collection of rewards, where a lump sum is paid to the spouse at a particular
age, and show how to use our results to derive the cashflow for this policy.

\begin{example}
\label{example:G810Liability}
Recall that we in Example \ref{example:G810} identified the payment functions for a G82 reward no. 810
as given by the formula (\ref{example:G810Payment}). We wish to calculate the liability of a reward no. 810
by applying Corollary \ref{corr:SpousePensionLiability} to this set of payment functions. To this end, we first need to
specify a distribution for $Z$. Here, let $Z^{u,y}$ be an inhomogeneous Markov chain on $[u-y,\infty)$ with two states $\{a,d\}$, initial state $a$ at time $u$ and
intensity $r\mapsto q_{ad}(r+y-u)$ for transitioning from $a$ to $d$ and zero intensity for
transitioning from $d$ to $a$. This intensity corresponds to having $q_{ad}(y)$ be the intensity for dying 
at age $y$. Now, it is immediate from (\ref{example:G810Payment}) that $\Pi_{y,u,u}(z)=0$ for all $y,u\ge0$.
Furthermore, note that for $y\neq \partial$, we have
\begin{align}
  E\Pi_{y,u,t}(Z^{u,y})
  &=1_{(t\ge u)}\int_u^t P(Z^{u,y}_r=a)\dv r,
\end{align}
so that $t\mapsto E\Pi_{y,u,t}(Z^{u,y})$ is differentiable for $t>u$, and the derivative is
\begin{align}
  \frac{\dv}{\dv t}E\Pi_{y,u,t}(Z^{u,y})
  &=P(Z^{u,y}_t=a)
  =\exp\left(\int_u^t q_{ad}(r+y-u)\dv r\right)\notag\\
  &=\exp\left(\int_y^{y+t-u} q_{ad}(r)\dv r\right),
\end{align}
corresponding to the survival probability of the spouse from age $y$ at time $u$ to age $y+t-u$ at
time $t$. As a consequence, we obtain from Corollary \ref{corr:SpousePensionLiability} that the liability
is given by
\begin{align}
 L  &= \int_0^\infty h(u)g(u) \int_0^\infty f(y|u) \int_u^\infty  e^{-rt} \exp\left(\int_y^{y+t-u} q_{ad}(r)\dv r\right) \dv t \dv y \dv u,
\end{align}
corresponding to e.g. the formula for the liability found in Section 9 of the G82 concession.
\hfill$\circ$
\end{example}

\begin{example}
\label{example:G815Longevity}
In this example, we show how our generic model includes a G82-type model for a terminating spouse's
pension, corresponding to reward no. 815, including the application of longevity improvements when
calculating the value of the terminating annuity for the spouse. A terminating spouse's pension
pays an annuity upon the death of the insured, conditional on the spouse having an age less than
a given termination age $c$. As in Example \ref{example:G810}, let $E=\{a,d\}$. Define, for $t\ge 0$,
\begin{align}
\label{example:G815Payment}
  \Pi_{y,u,t}(z) &= 1_{(y\neq\partial)}1_{(t\ge u)}\int_u^t 1_{(z_r=a)}1_{(y+r-u\le c)}\dv r.
\end{align}
The payment functions $(\Pi_{y,u})$ then correspond to the payments of a terminating spouse's pension
with termination age $c$. Next, assume given e.g. a nonnegative and continuous mapping $q:\RR_+\times\RR_+\to\RR_+$,
where $q(t,y)$ represents the intensity for dying at time $t$ when the age of the spouse is $y$ at time $t$. Assume
that $Z^{y,u}$ is an inhomogeneous Markov chain on $[u-y,\infty)$ with state space $E$, intensity
$r\mapsto q(r,y+r-u)$ for transitioning from $a$ to $d$ and initial state $a$. This corresponds to that when the spouse
has age $y$ at time $u$, the intensity for transitioning from $a$ to $d$ at time $r\ge u$ is $q(r,y+r-u)$. We
then have $\Pi_{y,u,u}(z)=0$ for all $y,u\ge0$, and for $y\neq \partial$, we have that 
$t\mapsto E\Pi_{y,u,t}(Z^{u,y})$ is differentiable for $t>u$, with derivative
\begin{align}
  \frac{\dv}{\dv t}E\Pi_{y,u,t}(Z^{u,y})
  &=P(Z^{u,y}_t=a)1_{(y+t-u\le c)}\notag\\
  &=\exp\left(\int_u^t q_{ad}(r,y+r-u)\dv r\right),
\end{align}
corresponding to the survival probability of the spouse from age $y$ at time $u$ to age $y+t-u$ at
time $t$. Corollary \ref{corr:SpousePensionCashflow} then yields that the cashflow of the reward is
\begin{align}
\label{eq:CashflowG815}
 a_t  &= \int_0^t h(u)g(u) \int_0^\infty f(y|u) \exp\left(\int_u^t q_{ad}(r,y+r-u)\dv r\right) \dv y\dv u,
\end{align}
corresponding to e.g. the formula for the liability found in Section 9 of the G82 concession,
with the exception of the intensity $q_{ad}$ depending not only on age but also on time.
\hfill$\circ$
\end{example}

\begin{example}
\label{example:LumpSumLongevity}
In this example, we show how our generic model is capable of modeling non-G82 type rewards, in
this case an insurance policy paying out a lump sum of one unit of monies to the spouse, when
the spouse reaches age $c$, if the spouse at that time remains alive. If the age of the spouse
at the death of the insured is $c$ or older, the lump sum is paid out immediately. To define the payment
functions, let $E=\{a,d\}$. Define, for $t\ge 0$,
\begin{align}
\label{eq:LumpsumExamplePayment}
  \Pi_{y,u,t}(z) &= 1_{(y\neq\partial)} 1_{(t\ge u)} 1_{(y+t-u\ge c)}1_{(z_{u+c-y}=a)}.
\end{align}
These payment functions corresponds to receiving one unit of monies when achieving the age $c$ (corresponding
to the indicator for $y+t-u\ge c)$ for any time $t\ge u$, or receiving one unit of monies at time $u$
if the age $c$ already is achieved at that time, with all payments contingent upon the spouse being alive at
age $c$. Note that as $c\ge0$ and $z\in D_{u-y}(E)$, the expression $z_{u+c-y}$ is well-defined. As in
Example \ref{example:G815Longevity}, assume given a continuous mapping $q:\RR_+\times\RR_+\to\RR_+$,
and assume that $Z^{y,u}$ is an inhomogeneous Markov chain on $[u-y,\infty)$ with state space $E$, intensity
$r\mapsto q(r,y+y-u)$ for transitioning from $a$ to $d$ and having state $a$ at time $u$. From
Theorem \ref{theorem:ECCFormula}, we know that the expected cumulative payments are
\begin{align}
\label{eq:LumpsumCumulativeCashflow}
  A_t  &=\int_0^t h(u)g(u) \int_0^\infty f(y|u) E\Pi_{u,y,t}(Z^{u,y}) \dv y\dv u.
\end{align}
We wish to identify the corresponding cashflow. In contrast to the previous examples, it does not generally hold in this case that $\Pi_{y,u,u}(z)=0$ for
all $y,u\ge0$, since lump sum payments can be made immediately upon the death of the insured. However, we do have that
for $y\neq\partial$ and $t\ge u$ that
\begin{align}
  E\Pi_{y,u,t}(Z^{u,y})
  &=1_{(y+t-u\ge c)} P(Z^{u,y}_{u+c-y}=a).
\end{align}
Here, if $c-y\le 0$, it holds with probability one that $Z^{u,y}_{u+c-y}=a$, since transitions from state
$d$ to $a$ are not possible and the state of $Z^{u,y}$ at time $u$ is $a$. Therefore, we obtain for $t\ge u$ that
\begin{align}
  E\Pi_{y,u,t}(Z^{u,y}) &=1_{(y+t-u\ge c)} = 1 \textrm{ for }c \le y, \label{eq:LumpSumExpected1}\\
  E\Pi_{y,u,t}(Z^{u,y}) &= 1_{(y+t-u\ge c)} \exp\left(\int_u^{u+c-y} q_{ad}(r,y+r-u)\dv r\right)\textrm{ for }c > y.\label{eq:LumpSumExpected2}
\end{align}
As a consequence, $E\Pi_{y,u,t}(Z^{u,y})$ is generally not differentiable as a function of $t$, since
it may contain a jump at time $c+u-y$, corresponding to the time when the spouse reaches age $c$. Now define
\begin{align}
  \xi(u,y) &= \exp\left(\int_u^{u+c-y} q_{ad}(r,y+r-u)\dv r\right) \dv y,
\end{align}
we then obtain for $0\le u\le t$ that
\begin{align}
  \int_0^\infty f(y|u) E\Pi_{u,y,t}(Z^{u,y}) \dv y
  &=\int_c^\infty f(y|u) \dv y+\int_0^c f(y|u) 1_{(t\ge u+c-y)} \xi(u,y) \dv y.\notag
\end{align}
As $u+c-t\le c$, we obtain that 
\begin{align}
  \frac{\dv}{\dv t}\int_0^c f(y|u) 1_{(t\ge u+c-y)} \xi(u,y) \dv y
  &=\frac{\dv}{\dv t}\int_{0\lor (u+c-t)}^c f(y|u) \xi(u,y) \dv y\notag\\
  &=1_{(u\ge t-c)}f(c+u-t|u)\xi(u,c+u-t),
\end{align}
where the derivative exists for $t\neq u+c$. Furthermore, (\ref{eq:LumpSumExpected1}) and (\ref{eq:LumpSumExpected2}) yields
\begin{align}
  \int_0^\infty f(y|u) E\Pi_{t,y,t}(Z^{t,y}) \dv y
  &=\int_c^\infty f(y|u)\dv y.
\end{align}
Therefore, the Leibniz integration rule allows us to conclude that the cashflow exists and is given by
\begin{align}
  a_t &= \int_0^t \frac{\dv}{\dv t}h(u)g(u) \int_0^\infty f(y|u) E\Pi_{u,y,t}(Z^{u,y}) \dv y\dv u\notag\\
      &+ h(t)g(t)\int_0^\infty f(y|u) E\Pi_{t,y,t}(Z^{t,y}) \dv y\notag\\
      &= \int_0^t h(u)g(u) 1_{(u\ge t-c)}f(c+u-t|u)\exp\left(\int_u^t q_{ad}(r,c+r-t)\dv r\right) \dv y\dv u\notag\\
      &+h(t)g(t)\int_c^\infty f(y|u)\dv y.
\end{align}
Here, the first term corresponds to the expected payments per time given that the insured dies as married at time $u\le t$.
For payments to occur at time $t$, the spouse must have age $c$ at time $t$, which corresponds to having age $c+u-t$ at
time $u$, which occurs with density $f(c+u-t|u)$, and payments then occur when the spouse survives to age $c$, which
corresponds to the exponential. The second term results from the immediate payments made at time $t$ when the spouse dies
at $t$, and these payments occur precisely when the spouse has age $c$ or more at time $t$, which corresponds to the
integral of $f(y|u)$ over $[c,\infty)$.
\hfill$\circ$
\end{example}

\section{An MPP model for marriage and spouse age with longevity}

\label{sec:LongevityMPP}

Recall that the generic model outlined in Section \ref{sec:generic} consists of variables $T$, $X$, $Y$
and a set of processes $Z^{u,y}$ for $u,y\ge0$. In the expression for e.g. the expected cumulative cashflows,
see (\ref{eq:CumulativeCashflow}), the expressions involved are $h(u)$, $g(u)$, $f(y|u)$ and $E\Pi_{u,y,t}(Z^{u,y})$. In
order to calculate cashflows and liabilities, it is necessary to model and calculate these four expressions. In this
section, we develop a marked point process model allowing us to express $g(u)$ and $f(y|u)$ in terms of intensities for marriage,
divorce and the death of the spouse.

Note that specification of a model yielding a credible expression for $h(u)$ is generally not a problem, as this
can be done by e.g. specifying a Markov chain model for the health state of the insured and letting $T$ be
the hitting time of the death state. Likewise, as we have seen in the examples of the previous section,
in order to obtain explicit expressions for $E\Pi_{u,y,t}(Z^{u,y})$, it suffices to specify e.g. a Markov
chain model for the health state of the spouse conditionally on the time of death of the insured and the
age of the insured at that time.

In order to obtain a full specification of all expressions necessary to
calculate expected cumulative cashflows et cetera, it therefore suffices to specify a model yielding expressions for
$g(u)$ and $f(y|u)$, which is the objective of this section. The motivations
for considering this particular marked point process model of this section are twofold: We aim to obtain both a
formalization of the model yielding the formulas of the G82 concession, as well as an extended model including longevity improvements for the spouse.

In Subection \ref{subsec:G82CollProcesses}, we construct the basic model framework, specifying
a probability space with variables $T$, $X$ and $Y$ depending on an underlying marked point
process. In Subsection \ref{subsec:MaritalExpr}, we derive expressions for $g(u)$ and $f(y|u)$
in this model. Finally, in Subsection \ref{subsec:G82Comparison}, we show how to obtain the
G82 expressions as a special case of our results.

\subsection{Model construction}

\label{subsec:G82CollProcesses}

In this subsection, we construct the marked point process based model for $T$, $X$ and $Y$ mentioned
above. Our model must be such that we can express and calculate $g(u)=P(X=x_m|T=u)$, the conditional probability of marriage given
the time $T$ of death of the insured, as well as the conditional density $f(y|u)$ of $Y$ given $T=u$ and $X=x_m$.
As modeling of $T$ is not our main interest, we will simply assume that a variable $T$ with density
$h$ is given. In order to obtain a joint model for $T$, $X$ and $Y$, it remains to specify $X$ and $Y$.
To this end, we will develop a marked point process model for the combined marital state of the
insured and the health state of the spouse, allowing for remarriage of the insured after the
death of a spouse. We will assume that the distribution of the marital state of the insured at the time of death is the
same as the distribution of the marital state at the time of death if the insured were not capable of dying. This
removes the need for explicitly modeling the health state of the insured jointly with the marital state
of the insured.

We let $Z^\mu_t$ denote the marital state of the insured at time $t$, with state space $E^\mu$ given by $E^\mu=\{s_0,m_1,s_1,m_2,s_2,\ldots,\}$. We think of the
state $s_i$ as corresponding to the $i$'th single (unmarried) state and think of $m_i$ as corresponding
to the $i$'th married state. For convenience, we let $s=\{s_0,s_1,s_2,\ldots\}$ and
$m=\{m_1,m_2,m_3,\ldots\}$. Furthermore, we let $Z^\zeta_t$ denote the health state of the current spouse
at time $t$, with state space $E^\zeta=\{a,d,\partial\}$, where, similarly to the previous section, $\partial$ is a coffin state, in this
case the state held by the process $Z^\zeta$ when the insured is unmarried, $a$ corresponds to the spouse
being alive and $d$ corresponds to the spouse being dead. Finally, we let $U_t^\zeta$ denote the age of
the spouse at time $t$, with state space $\RR_+\cup\{\partial\}$. We assume that $(Z^\mu,Z^\zeta,U^\zeta)$
is independent of $T$.

\begin{figure}[htb]
\begin{center}
	\begin{tikzpicture}[node distance=0.8cm and 0.8cm]
	%Node declaration.
	\node[markovnodeSmall] (node1) {$(s_0,\partial)$};
	\node[markovnodeSmall] (node2) [right = of node1] {$(m_1,a)$};
	\node[markovnodeSmall] (node3) [above right = of node2] {$(s_1,a)$};
	\node[markovnodeSmall] (node4) [below right = of node2] {$(s_1,d)$};
	\node[markovnodeSmall] (node5) [right = of node2] {$(m_2,a)$};
	\node[markovnodeSmall] (node6) [above right = of node5] {$(s_2,a)$};
	\node[markovnodeSmall] (node7) [below right = of node5] {$(s_2,d)$};
	\node[markovnodeSmall] (node8) [right = of node5] {$(m_3,a)$};
	\node[markovnodeSmall] (node9) [above right = of node8,draw=none] {};
	\node[markovnodeSmall] (node10) [below right = of node8,draw=none] {};
		%Arrow declaration.
	\path (node1) edge[markovarrow] (node2);
	\path (node2) edge[markovarrow] (node3);
	\path (node2) edge[markovarrow] (node4);
	\path (node3) edge[markovarrow] (node5);
	\path (node4) edge[markovarrow] (node5);
	\path (node5) edge[markovarrow] (node6);
	\path (node5) edge[markovarrow] (node7);
	\path (node6) edge[markovarrow] (node8);
	\path (node7) edge[markovarrow] (node8);
	\path (node8) edge[markovarrowdashed] (node9);
	\path (node8) edge[markovarrowdashed] (node10);
	\end{tikzpicture}
\end{center}
\caption{Transition graph of the combined marked point process for the marital state $Z^\mu$ of
	the insured and the health state $Z^\zeta$ of the spouse.}
\label{figure:G82CollectiveCombinedProcess}
\end{figure}
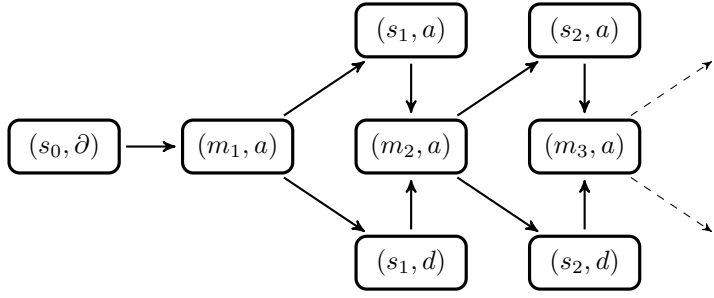

Our next task is to construct and specify the joint distribution of the processes $(Z^\mu,Z^\zeta,U^\zeta)$ in
the model. To this end, first define stopping times
\begin{align}
  T^\mu_i &= \inf\{t\ge0 \mid Z^\mu_t = i\}
\end{align}
for any $i\in E^\mu$. Assume given an uncountable sequence of variables $(Y^{\zeta,t})_{t\ge0}$, all independent.
We interpret $Y^{\zeta,t}$ as the hypothetical initial age of a spouse married at age $t$. We assume
that the variables $Y^\zeta_t$ are such that $Y^\zeta_t$ has density $\varphi(\cdot|t)$. We then define
\begin{align}
  U^{\zeta,\nu}_t &= Y^\zeta_{T^\mu_{m_\nu}}+(t - T^\mu_{m_\nu}) \label{eq:Uzetanuxdef}\\
  U^\zeta_t &= U^{\zeta,\nu}_t \textrm{ when } (Z^\mu_t,Z^\zeta_t) \in\{(m_\nu,a),(s_\nu,a)\}
                               \textrm{ and } U^\zeta_t =\partial \textrm{ otherwise}.
\end{align}
Note that in (\ref{eq:Uzetanuxdef}), the process $U^{\zeta,\nu}_x$ increases indefinitely
after $T^\mu_{m_\nu}$. Consistently with this, $U^{\zeta,\nu}_x$ is the hypothetical age of the $\nu$'th spouse
given that the spouse cannot die. We then let $(Z^\mu,Z^\zeta)$ be a marked point process with state space $E^\mu\times E^\zeta$,
initial state $(s,\partial)$ and with intensity $\lambda:\RR_+\times (E^\mu\times E^\zeta)\times (E^\mu\times E^\zeta)$ given by
\begin{align}
  \lambda(t,(s_{\nu-1},i),(m_{\nu},a)) &= \gamma(t) \textrm{ for all } i\in E^\zeta,\\
  \lambda(t,(m_\nu,a),(s_{\nu},a)) &= \sigma(t), \\
  \lambda(t,(m_\nu,a),(s_{\nu},d)) &= q^\zeta(t,U^\zeta_t),
\end{align}
and all other intensities zero. For details on marked point processes and intensities, see \cite{MR2189574}.  Here, $\gamma:\RR_+\mapsto\RR_+$ denotes the marriage intensity,
$\sigma:\RR_+\mapsto\RR_+$ denotes the divorce intensity and $q^\zeta:\RR_+\times\RR_+\mapsto\RR_+$ denotes the death
intensity for the spouse, with $q^\zeta(t,y)$ denoting the intensity at time $t$ when the spouse has age $y$ at time $t$. Note
in particular that the states $(s_0,a)$ and $(s_0,d)$ never occur,
corresponding to the impossibility of having a spouse in the initial single state $s_0$. See
Figure \ref{figure:G82CollectiveCombinedProcess} for an illustration of the possible transitions of the process.

We have now made sufficient assumptions on the processes $Z^\mu$ ,$Z^\zeta$ and $U^\zeta$ to uniquely
determine their joint distribution. We then simply define
\begin{align}
  X &= \left\{\begin{array}{ll}
        x_s & \textrm{ when } Z^\mu_T \in s \\
        x_m & \textrm{ when } Z^\mu_T \in m
        \end{array}\right. \\
  Y &= U^\zeta_T.
\end{align}
This yields a complete model for $T$, $X$ and $Y$, and we may then express $g(u)$ and $f(y|u)$
in terms of the underlying marked point process. To see how this can be done, note that by
the independence of $T$ and $(Z^\mu,Z^\zeta,U^\zeta)$, we have
\begin{align}
  g(u) &= P(X = x_m | T = u)
        = P(Z^\mu_T \in m | T = u)
        = P(Z^\mu_u \in m).
\end{align}
Furthermore, for measurable $A\subseteq\RR_+$, we have
\begin{align}
  \int_A f(y|u)\dv y &= P(Y \in A |T = u, X = X_m)
        = P(U^\zeta_T \in A |T = u, Z^\mu_T \in m)\notag\\
        &=P(U^\zeta_u \in A |T = u, Z^\mu_u \in m)
         =P(U^\zeta_u \in A |Z^\mu_u \in m).
\end{align}
Since $A$ was arbitrary, this yields that $f(\cdot|u)$ is the density of $U^\zeta_u$ given $Z^\mu_u \in m$,
assuming that this density exists. In conclusion, we now have that
\begin{align}
  g(u) &= P(Z^\mu_u \in m) \label{eq:guExpression}\\
  f(\cdot|u) &= \frac{\dv}{\dv m_\ell}P(U^\zeta_u \in \cdot |Z^\mu_u \in m),\label{eq:fyuExpression}
\end{align}
where $m_\ell$ denotes the Lebesgue measure and the derivative refers to the Radon-Nikodym derivative.
Our next objective is to obtain expressions for (\ref{eq:guExpression}) and (\ref{eq:fyuExpression}).

\subsection{Expressions for the marriage probability and spouse's age density}

\label{subsec:MaritalExpr}

In this subsection, we state expressions for (\ref{eq:guExpression}) and (\ref{eq:fyuExpression}) in terms
of the intensities for marriage, divorce and spouse's death. To
do so, we first introduce the auxiliary expressions
\begin{align}
	u_\nu(t) &= P(Z^\mu_t = s_\nu) \textrm{ for }\nu=0,1,\ldots,\label{eq:unuxDef}\\
	g_\nu(\cdot|t) &= \frac{\dv}{\dv m_\ell} P(U^\zeta_t\in \cdot,Z^\mu_t = m_\nu) \textrm{ for }\nu=1,2,\ldots,\label{eq:gnuetaxDef}
\end{align}
and further define
\begin{align}
\ell^\gamma_t &= \exp\left(-\int_0^t \gamma(v)\dv v\right), \\
\ell^\sigma_t &= \exp\left(-\int_0^t \sigma(v)\dv v\right), \\
\ell^\zeta_{t,y} &= \exp\left(-\int_0^t q^\zeta(v,y+v-t)\dv v\right).\label{eq:SpouseSurvivalLongevity}
\end{align}
The interpretation of the latter expression is that $\ell^\zeta_{t,y}$ is the survival probability
for a spouse starting at time zero and until time $t$, with the assumption that the spouse,
if surviving, will be $y$ years at age $t$. In Theorem \ref{theorem:gxExpression}
and Theorem \ref{theorem:fetaxExpression}, we state results on how to express (\ref{eq:guExpression}) and
(\ref{eq:fyuExpression}) in terms of these auxiliary expressions. Before stating the two theorems, we first state
three lemmas, yielding expressions for (\ref{eq:unuxDef}) and (\ref{eq:gnuetaxDef}).

\begin{lemma}
\label{lemma:u0xSource}
It holds that $u_0(t) = \ell^\gamma_t$ for $t\ge 0$.
\end{lemma}

The following lemma shows that the density defined in (\ref{eq:gnuetaxDef}) in
fact exists, and yields an expression for the density. In the following, we let $\BBB_+$ denote the
Borel $\sigma$-algebra on $\RR_+$.

\begin{lemma}
\label{lemma:gnuetaxSource}
Let $\nu\ge1$ and define a measure $Q_{t,\nu}:\BBB_+\to[0,1]$ by
\begin{align}
\label{eq:gnuetaxMeasure}
  Q_{t,\nu}(A) &= P(U^\zeta_t \in A, Z^\mu_t = m_\nu).
\end{align}
Then $Q_{t,\nu}$ has a density $g_\nu(\cdot|t)$ with respect to the Lebesgue measure, and the
density is given by
\begin{align}
\label{eq:gnuetatDensity}
  g_\nu(y|t)
  &=\int_0^t u_{\nu-1}(v)\gamma(v) \varphi(y+v-t|v)\frac{\ell^\sigma_t}{\ell^\sigma_v}\frac{\ell^\zeta_{t,y}}{\ell^\zeta_{v,y+v-t}} \dv v
\end{align}
\end{lemma}

\begin{lemma}
\label{lemma:unuxSource}
For $\nu\ge1$, it holds that
\begin{align}
\label{eq:unuExpr}
  u_\nu(t) &= \int_0^\infty \int_0^t g(y|v) (\sigma(v)+ q^\zeta(v,y)) \frac{\ell^\gamma_t}{\ell^\gamma_v} \dv v\dv y.
\end{align}
\end{lemma}

Note that combining Lemma \ref{lemma:u0xSource}, Lemma \ref{lemma:gnuetaxSource}
and Lemma \ref{lemma:unuxSource}, we obtain expressions for $u_\nu(x)$ and $g_\nu(\eta|x)$
in terms of coupled recursion equations. The interpretation of the above lemmas is as follows. As
regards Lemma \ref{lemma:u0xSource}, this lemma simply states that $u_0(t)$ is the survival probability
for the distribution with hazard $\gamma$, corresponding to that the only way of leaving the first
single state $s_0$ is to become married with intensity $\gamma$. The formula (\ref{eq:gnuetatDensity})
expresses that the density of being married for the $\nu$'th time at time $t$ with a spouse of age $y$ at time $t$ can be obtained
by, for some $v\le t$, being single for the $\nu-1$'th time at time $v$, becoming married with intensity
$\gamma(v)$, to a spouse which at time $t$ has age $y$, corresponding to having age $y+v-t$ at time $v$,
and finally, not being divorced in the time interval $[v,t]$ with intensity $\sigma$ and not having the
spouse die in the time interval $[v,t]$ with intensity $r\mapsto q^\zeta(r,y+r-t)$. Similarly, the formula
(\ref{eq:unuExpr}) expresses that the probability of being single for the $\nu$'th time at time $t$, for
$\nu\ge1$, can be obtained by conditioning on being married for the $\nu$'the time at time $v\le t$ with a spouse whose age at time $v$ is $y$,
occurring with density $g_\nu(y|v)$, and afterwards either being divorced with intensity $\sigma(v)$ or
having the spouse die with intensity $q^\zeta(v,y)$, and finally not remarrying, where marriage has intensity $\gamma$.

With these lemmas at hand, we are now ready to state our main results yielding expressions for $g(t)$ and $f(y|t)$.

\begin{theorem}
\label{theorem:gxExpression}
With $g(t) = P(Z^\mu_t \in m)$ as in (\ref{eq:guExpression}), it holds that
\begin{align}
\label{eq:gxExpression}
  g(t) &= \sum_{\nu=1}^\infty \int_0^\infty g_\nu(y|t)\dv y.
\end{align}
\end{theorem}

\begin{theorem}
\label{theorem:fetaxExpression}
Define $Q_t:\BBB_+\mapsto[0,1]$ by $Q_t(A) = P(U^\zeta_t \in A|Z^\mu_t =m)$. It then
holds that $Q_t$ has a density with respect to the Lebesgue measure. With $f(\cdot|t)$
denoting the density as in (\ref{eq:fyuExpression}), the density is
\begin{align}
\label{eq:fetaxExpression}
  f(y|t) &= \frac{1}{g(t)}\sum_{\nu=1}^\infty g_\nu(y|t).
\end{align}
\end{theorem}

Theorem \ref{theorem:gxExpression} and Theorem \ref{theorem:fetaxExpression} expresses $g(t)$ and $f(y|t)$ in
terms of the auxiliary density $g_\nu(y|t)$, which in turn is obtained from the fundamental model parameters
and $u_\nu(t)$. Essentially, (\ref{eq:gxExpression}) states that the probability of being married can be
obtained by integration of the density corresponding to being married with a spouse of a particular age,
and (\ref{eq:fetaxExpression}) states that the density of the spouse's age given marriage can be obtained
by considering the density of the measure corresponding to being married with a spouse of a particular
age and normalizing with the probability of being married in order to obtain the conditional density.

\subsection{Comparison to the G82 model}

\label{subsec:G82Comparison}

In this subsection, we show how to reclaim the expressions for the marriage probability and spouse
age density from the G82 model, see Section 8 of the G82 concession. To this end, consider the
model setup of Subsection \ref{subsec:G82CollProcesses}, and assume that the insured
at time $t$ has age zero. We further assume no longevity improvements for the spouse,
meaning that we let $q^\zeta(t,y) = q^\zeta(y)$, with minor abuse of notation, corresponding
to $q^\zeta(y)$ denoting the intensity for the spouse's death at age $y$. Furthermore, we
assume that $\sigma(t)=\gamma(t)=0$ for $t\le a$, corresponding to a lower age bound for marriage.
In order to obtain consistency with the G82 formulas, we let $x$ denote time in the following,
or equivalently, the age of the insured, and we let $\eta$ denote a generic value of the age
of the spouse. Furthermore, we extend $g_\nu(\eta|x)$ and $f(\eta|x)$ from $\RR_+$ to $\RR$ by
letting $g_\nu(\eta|x) = f(\eta|x)=0$ for $\eta\le0$. 

With this notation, Theorem \ref{theorem:gxExpression} and Theorem \ref{theorem:fetaxExpression} yields
\begin{align}
  g(x) &= \sum_{\nu=1}^\infty \int_{-\infty}^\infty g_\nu(\eta|x)\dv \eta,\\
  f(\eta|x) &= \frac{1}{g(x)}\sum_{\nu=1}^\infty g_\nu(\eta|x).
\end{align}
Now define
\begin{align}
  \ell^\zeta_\eta &= \exp\left(-\int_0^\eta q^\zeta(y)\dv y\right).
\end{align}
Recalling (\ref{eq:SpouseSurvivalLongevity}), we then have 
\begin{align}
  \frac{\ell^\zeta_{x,\eta}}{\ell^\zeta_{\xi,\eta+\xi-x}}
  &=\exp\left(\int_\xi^x q^\zeta(v,\eta+v-x)\dv v\right)\notag\\
  &=\exp\left(\int_\xi^x q^\zeta(\eta+v-x)\dv v\right)
   =\frac{\ell^\zeta_\eta}{\ell^\zeta_{\eta+\xi-x}}.
\end{align}
Therefore, Lemma \ref{lemma:gnuetaxSource} and our assumption that $\gamma(\xi)=0$ for $\xi\le a$ yields
\begin{align}
  g_\nu(\eta|x)
  &=\int_0^x u_{\nu-1}(\xi)\gamma(\xi) \varphi(\eta+\xi-x|\xi)\frac{\ell^\sigma_x}{\ell^\sigma_\xi}\frac{\ell^\zeta_{x,\eta}}{\ell^\zeta_{\xi,\eta+\xi-x}} \dv \xi\notag\\
  &=\int_a^x u_{\nu-1}(\xi)\gamma(\xi) \varphi(\xi+\eta-x|\xi)\frac{\ell^\sigma_x}{\ell^\sigma_\xi}\frac{\ell^\zeta_{\eta}}{\ell^\zeta_{\eta+\xi-x}} \dv \xi.
\end{align}
Furthermore, Lemma \ref{lemma:unuxSource} yields
\begin{align}
  u_\nu(x) &= \int_{-\infty}^\infty \int_0^x g(y|\xi) (\sigma(\xi)+ q^\zeta(y)) \frac{\ell^\gamma_x}{\ell^\gamma_\xi} \dv \xi\dv y\notag\\
  &= \int_{-\infty}^\infty \int_0^x g(\xi+\eta-x|\xi) (\sigma(\xi)+ q^\zeta(\xi+\eta-x)) \frac{\ell^\gamma_x}{\ell^\gamma_\xi} \dv \xi\dv \eta.
\end{align}
Here, the change of variable from $y$ to $\xi+\eta-x$ corresponds to a change of variable from the spouse age $y$ when the insured
has age $\xi$ to the spouse age $\eta$ when the insured has age $x$. Finally, the assumption that $\gamma(\xi)=0$
for $\xi\le a$ immediately yields that $u_0(x)= 1$ for $x\le a$ and $u_0(x)= \ell^\gamma_x / \ell^\gamma_a$ for $x\ge a$. All
the above formulas now match exactly the formulas from Section 8 of the G82 concession, showing how to obtain these
formulas as a special case of our generic model in general and our marked point process model in particular.

\section{Discussion}

\label{sec:Discussion}

In this paper, we have introduced a generic model for spouse's pensions, and have derived expressions
for expected cumulative payments, cashflows and liabilities. By example, we have shown that these
results can be used for the calculation of the cashflows and liabilities for both classical G82-type
spouse's pensions as well as for more advanced policies such as the policy considered in e g.
Example \ref{example:LumpSumLongevity}. Furthermore, we have developed an
explicit joint marked point process model for the marital state of the insured and the health state of the spouse, including
longevity improvements for the spouse, and have shown how to derive the expressions relevant for
cashflows et cetera using this model. Finally, we have shown that in the absence of longevity
effects for the spouse's health state, the expressions for the marital probability and the density of the spouse's age from
the G82 concession can be reclaimed from the results obtained here.

The immediate benefit of the generic model developed in Section \ref{sec:generic} is the
result that common expressions for cashflows and liabilities hold independently of the particular
model for marital probabilities and spouse's ages. This allows for a simple framework for deriving
liabilities for various types of spouse's pensions without consideration of the particular model for
marital behaviour to be applied.

The results also makes explicit that one main issue for the calculation of liabilities for spouse's
pensions is the estimation of the probability of marriage $g(t)$
and the density of the spouse's age $f(y|t)$. One opportunity for further work is to consider
methods for efficient estimation of these functions, e.g. through direct estimation from
the marital states of a general population, or through estimation of intensities in a more
specific model such as the marked point process model considered in Section \ref{sec:LongevityMPP}.

The results of Section \ref{sec:LongevityMPP} opens up for the possibility of obtaining more accurate
estimates of liabilities by taking into account the longevity improvements of the spouse when calculating
the probability of marriage for the insured. We take a moment to reflect upon the complications of
numerical computations resulting from this extension. In the classical G82 model, one component
of the numerical effort required for computations is the evaluation of $g(x)$ and $f(\eta|x)$, see
Subsection \ref{subsec:G82Comparison}, for $x=0,\ldots,125$ and $\eta=0,\ldots,125$, with 125 conventionally
being taken in the G82 concession as the maximal age of the insured and the spouse. In Subsection
\ref{subsec:MaritalExpr}, these expressions are parameterized in terms of time, i. e. $g(t)$ and $f(y|t)$,
and depend on given intensities $\gamma$ and $\sigma$ for marriage and divorce, respectively. For
the liabilities corresponding to a single insured, therefore, the computational effort required is
no greater than in the case without longevity. For an entire population, however, some computational
overhead occurs: The intensities $\gamma$ and $\sigma$ correspond to marriage and divorce intensities
as a function of time. Given a model where these intensities ultimately depend on the age of the
insured, we would have e.g. $\gamma(t) = \gamma_a(x_0 + t)$ and $\sigma(t)=\sigma_a(x_0+t)$, where
$\gamma_a$ and $\sigma_a$ denote the intensities for marriage and divorce as functions of the age
of the insured, and $x_0$ denotes the age at time zero of the insured. As a result, the marital expressions
$g(t)$ and $f(y|t)$ would in fact depend on the initial age of the insured. For the calculation of
the total liabilities for a pension fund, it would therefore generally be necessary to calculate
a grid of values $g(x,x_0)$ and $f(\eta|x,x_0)$ for $x_0=0,\ldots,125$, $x\ge x_0$ and $y=0,\ldots,125$. Other than
this, no particular increase in computational complexity would be incurred from the inclusion
of longevity improvements.

As regards opportunities for further work, the most pressing necessity for the accurate estimation
of liabilities is the estimation of $g(t)$ and $f(y|t)$. The issue of obtaining standard
methodology for this, either by direct estimation or by use of more complex models such as the
marked point process model, is not very well developed. Furthermore, as the calculation of
$g(t)$ and $f(y|t)$ in the marked point process model is relatively computationally intensive,
it is of interest to develop models for the marital behavior of the insured and the spouse which
is both amenable to estimation of parameters and in which simple expressions for $g(t)$ and
$f(y|t)$ can be obtained.

\appendix

\section{Proofs}

\label{sec:Proofs}

This appendix contains proofs of the results of the main part of the paper.

\subsection{Proofs of results in Section \ref{sec:generic}} In this subsection, we prove
the results of Section \ref{sec:generic} on expected cumulative payments, cashflows
and liabilities in the generic model for spouse's pensions.

\textbf{Proof of Theorem \ref{theorem:ECCFormula}.}
Fix $t\ge0$ and let $\Psi_t:\RR_+\times \{x_s,x_m\}\times \fvsp\to\RR$ be defined by
\begin{align}
  \Psi_t(u,x,F)
  &=\int_0^t 1_{(s\ge u)} 1_{(x = x_m)} \dv F_s.
\end{align}
Recalling (\ref{eq:PensionCPP}), it then holds that $B_t = \Psi_t(T,X,C)$. With $R$ denoting the distribution of
$(T,X)$ and $(Q_{u,x})$ denoting the conditional distribution of $C$ given $(T,X_m)$,
we then obtain that
\begin{align}
  EB_t &= E\Psi_t(T,X,C)
        = \int_{(\RR_+\cup\{\partial\})\times\{x_s,x_m\}} \int_{\fvsp}\Psi_t(u,x,F) \dv Q_{u,x}(F) \dv R(u,x)\notag\\
        &=\int_{[0,t]\times\{x_m\}} \int_{\fvsp}\int_0^t 1_{(s\ge u)} 1_{(x = x_m)} \dv F_s \dv Q_{u,x}(F) \dv R(u,x)\notag\\
        &=\int_{[0,t]\times\{x_m\}} \int_{\fvsp} F(t)-F(u-) \dv Q_{u,x}(F) \dv R(u,x).
\end{align}
Now note that
\begin{align}
  R([0,t]\times \{x_m\})
  &=P(T\le t,X = x_m)\notag\\
  &=\int_0^t h(u) P(X = x_m|T=u) \dv u = \int_0^t h(u)g(u)\dv u,
\end{align}
so that with $\pi:\RR_+\times\{x_m\}\to\RR_+$ defining the projection mapping onto the
first coordinate, it holds that the pushforward measure of the restriction of $R$ to $\RR_+\times\{x_m\}$ under $\pi$
has density $u\mapsto h(u)g(u)$ with respect to the Lebesgue measure. Inserting this into the above yields
\begin{align}
  EB_t
  &=\int_0^t h(u)g(u) \int_{\fvsp} F(t)-F(u-) \dv Q_{u,x_m}(F)\dv u.
\end{align}
Furthermore, by our assumptions, the conditional distribution of the variable $Y$ given
$T=u$ and $X=x_m$ has density $f(\cdot|u)$. Therefore, the conditional distribution
of $(T,Y)$ given $T=u$ and $X=x_m$ is the tensor product of the Dirac measure in $u$
and the measure with Lebesgue density $f(\cdot|u)$. With $Q_{u,x,y}$ denoting the conditional
distribution of $C$ given $T=u$, $X = x$ and $Y=y$, we therefore obtain for measurable $A\subseteq\fvsp$ that
\begin{align}
  Q_{u,x_m}(A)
  &=\int_0^\infty Q_{u,x_m,y}(A) f(y|u)\dv y.
\end{align}
Next, recalling that $C_t = \Pi_{T,Y,t}(Z^{T,Y})$, and further recalling that $Z^{u,y}$ for
all $u,y\ge0$ is assumed to be independent of $(T,X,Y)$, we obtain that the conditional distribution $Q_{u,x_m,y}$
is equal to the distribution of the process $t\mapsto \Pi_{u,y,t}(Z^{u,y})$, we obtain
\begin{align}
  &\int_{\fvsp} F(t)-F(u-) \dv Q_{u,x_m}(F)\notag\\
  &=\int_0^\infty f(y|u) \int_{\fvsp} F(t)-F(u-) \dv Q_{u,x_m,y}(F) \dv y\notag\\
  &=\int_0^\infty f(y|u) E((\Pi_{u,y,t}(Z^{u,y})-\Pi_{u,y,u-}(Z^{u,y}))  \dv y.
\end{align}
Collecting our conclusions and recalling that $\Pi_{u,y,u-}(z)=0$ for all $z\in D_u(E)$, we finally obtain
\begin{align}
  EB_t  &=\int_0^t h(u)g(u) \int_0^\infty f(y|u) E \Pi_{u,y,t}(Z^{u,y}) \dv y\dv u,
\end{align}
as required.
\hfill$\Box$

\textbf{Proof of Corollary \ref{corr:SpousePensionCashflow}.}
By Theorem \ref{theorem:ECCFormula}, we have
\begin{align}
 EB_t  &=\int_0^t h(u)g(u) \int_0^\infty f(y|u) E \Pi_{y,u,t}(Z^{u,y}) \dv y\dv u.
\end{align}
Applying the Leibniz integral rule, we then obtain
\begin{align}
 a_t &= \frac{\dv}{\dv t} EB_t = \int_0^t \frac{\dv}{\dv t} h(u)g(u) \int_0^\infty f(y|u) E \Pi_{y,u,t}(Z^{u,y}) \dv y\dv u\label{eq:LeibnizCFResult}\\
 &+ h(t)g(t) \int_0^\infty f(y|t) E \Pi_{y,t,t}(Z^{u,y}) \dv y.
\end{align}
By our assumptions, the latter term is zero. Next, as $y\mapsto f(y,u)$ is a probability density,
any bounded interval is integrable with respect to the measure with Lebesgue density $f(y|u)$. As
we have assumed that $E\Pi_{y,u,t}(Z^{u,y})$ is bounded as a function of $t$ on compact intervals, this yields that differentiation
under the inner integral in (\ref{eq:LeibnizCFResult}) is allowed, and we obtain the result
stated in the corollary.
\hfill$\Box$

\textbf{Proof of Corollary \ref{corr:SpousePensionLiability}.}
By approximation with Riemann sums and a uniform integrability argument, we have
\begin{align}
  L &= E \int_0^\infty e^{-rt} \dv B_t = \int_0^\infty e^{rt} \dv E B_t.
\end{align}
Applying Corollary \ref{corr:SpousePensionCashflow}, we then obtain
\begin{align}
  L 
   &= \int_0^\infty e^{-rt} \int_0^t h(u)g(u) \int_0^\infty f(y|u) \left(\frac{\dv}{\dv t} E \Pi_{y,u,t}(Z)\right) \dv y\dv u \dv t\notag\\
   &= \int_0^\infty h(u)g(u) \int_0^\infty f(y|u) \int_u^\infty  e^{-rt} \left(\frac{\dv}{\dv t} E \Pi_{y,u,t}(Z)\right) \dv t \dv y \dv u,
\end{align}
as required.
\hfill$\Box$

\subsection{Proofs of results in Section \ref{sec:LongevityMPP}} In this subsection, we prove
the results of Section \ref{sec:LongevityMPP} on expressions for the marriage probability
and the density of the spouse's age.

\textbf{Proof of Lemma \ref{lemma:u0xSource}.} This follows as 
\begin{align}
u_0(t) &= P(Z^\mu_t = s_0)
= P(Z^\mu_t = s_0,Z^\zeta_t = \partial)= \exp\left(-\int_0^t \gamma(v)\dv v\right),
\end{align}
since the intensity for leaving state $(s_0,\partial)$ is $\gamma$.
\hfill$\Box$

For the following lemmas, we require some results on compensators and intensities. For general
results on compensators, see Chapter V of \cite{MR1219534}. For results on compensators and
intensities in the particular context of marked point processes, see Chapter 3 of \cite{MR2189574}.

\begin{lemma}
\label{lemma:MaritalStoppingTime}
It holds that $T^\mu_{m_\nu}$ has a density $h_{m_\nu}:\RR_+\to\RR_+$ with respect to the
Lebesgue measure, and the density is given by $h_{m_\nu}(t) = P(Z^\mu_t = s_{\nu-1})\gamma(t)$.
\end{lemma}

\textbf{Proof of Lemma \ref{lemma:MaritalStoppingTime}.} Define $N^{m_\nu}$ by
\begin{align}
  N^{m_\nu}_t = \sum_{0<v\le t}1_{((Z^\mu_{v-}\neq m_\nu,Z^\mu_v=m_\nu)}.
\end{align}
It then holds that the compensator $A^{m_\nu}$ of
$N^{m_\nu}$ is
\begin{align}
  A^{m_\nu}_t &= \int_0^t \gamma(v) 1_{(Z^\mu_v = s_{\nu-1})}\dv v.
\end{align}
Since only a single jump to $m_\nu$ is possible, this yields
\begin{align}
  P(T^\nu_{m_\nu}\le t)
   &= EN^{m_\nu}_t
    = EA^{m_\nu}_t
   = E\int_0^t \gamma(v) 1_{(Z^\mu_v = s_{\nu-1})}\dv v\notag\\
   &= \int_0^t P(Z^\mu_v = s_{\nu-1}) \gamma(v) \dv v,
\end{align}
proving the lemma.
\hfill$\Box$

\textbf{Proof of Lemma \ref{lemma:gnuetaxSource}.}
Applying Lemma \ref{lemma:MaritalStoppingTime} and $(Z^\mu_t = m_\nu)\subseteq(T^\mu_{m_\nu}\le t)$, we obtain
\begin{align}
 P(U^\zeta_t \in A, Z^\mu_t = m_\nu)
&=P(U^\zeta_t \in A, Z^\mu_t = m_\nu, T^\mu_{m_\nu}\le t)\notag\\
&=\int_0^t P(U^\zeta_t \in A, Z^\mu_t = m_\nu|T^\mu_{m_\nu}=v) P(Z^\mu_v = s_{\nu-1})\gamma(v)\dv v\notag\\
&=\int_0^t u_{\nu-1}(v)\gamma(v) P(U^\zeta_t \in A, Z^\mu_t = m_\nu|T^\mu_{m_\nu}=v)\dv v.\label{eq:gnuetaxFirstRes}
\end{align}
The intensity of $Z^\mu$ leaving state $m_\nu$ at time $u$ is $\sigma(u)+q^\zeta(u,U^\zeta_u)$. Therefore, we obtain
\begin{align}
  &P(U^\zeta_t \in A, Z^\mu_t = m_\nu|T^\mu_{m_\nu}=v)\notag\\
  &=P(Y^\zeta_v+(t-v) \in A, Z^\mu_t = m_\nu|T^\mu_{m_\nu}=v)\notag\\
  &=\int_A \varphi(y+v-t|v)\frac{\ell^\sigma_t}{\ell^\sigma_v}\frac{\ell^\zeta_{t,y}}{\ell^\zeta_{v,y+v-t}}\dv y.
\end{align}
From this, we obtain that $P(U^\zeta_t\in\cdot, Z^\mu_t = m_\nu|T^\mu_{m_\nu}=v)$ has a density given by
\begin{align}
  y\mapsto \varphi(y+v-t|v)\frac{\ell^\sigma_t}{\ell^\sigma_v}\frac{\ell^\zeta_{t,y}}{\ell^\zeta_{v,y+v-t}}.\label{eq:gnuetaxSecondRes}
\end{align}
Combining (\ref{eq:gnuetaxFirstRes}) and (\ref{eq:gnuetaxSecondRes}), we find that $Q_{t,\nu}$ has a density,
and the density is
\begin{align}
  g_\nu(y|t)
  &=\frac{\dv}{\dv y} \int_0^t u_{\nu-1}(v)\gamma(v) P(U^\zeta_t \le y, Z^\mu_t = m_\nu|T^\mu_{m_\nu}=v)\dv v\notag\\
  &=\int_0^t u_{\nu-1}(v)\gamma(v) \frac{\dv}{\dv y}  P(U^\zeta_t \le y, Z^\mu_t = m_\nu|T^\mu_{m_\nu}=v)\dv v\notag\\
  &=\int_0^t u_{\nu-1}(v)\gamma(v) \varphi(y+v-t|v)\frac{\ell^\sigma_t}{\ell^\sigma_v}\frac{\ell^\zeta_{t,y}}{\ell^\zeta_{v,y+v-t}} \dv v,
\end{align}
as required.
\hfill$\Box$

\begin{lemma}
\label{lemma:SingleStoppingTime}
Let $\nu\ge1$. It then holds that $T^\mu_{s_\nu}$ has a
density $h_{s_\nu}:\RR_+\to\RR_+$ with respect to the Lebesgue measure, and the density is given by
\begin{align}
\label{eq:SingleStoppingTimeDensity}
  h_{s_\nu}(t) = \int_0^\infty g_\nu(y|t)(\sigma(t)+q^\zeta(t,y))\dv y
\end{align}
\end{lemma}

\textbf{Proof of Lemma \ref{lemma:SingleStoppingTime}.} Define a process $N^{s_\nu}$ by putting
\begin{align}
N^{s_\nu}_x = \sum_{0<y\le x}1_{((Z^\mu_{x-}\neq s_\nu,Z^\mu_x=s_\nu)}.
\end{align}
It then holds that the compensator $A^{s_\nu}$ of $N^{s_\nu}$ is
\begin{align}
A^{s_\nu}_t &= \int_0^t 1_{(Z^\mu_v = m_\nu)}(\sigma(v)+q^\zeta(v,U^\zeta_v))\dv v,
\end{align}
which yields
\begin{align}
P(T^\mu_{s_\nu}\le t) &=\int_0^t E1_{(Z^\mu_v = m_\nu)}(\sigma(v)+q^\zeta(v,U^\zeta_v))\dv v,
\end{align}
so that $T^\mu_{s_\nu}$ has density given by
$h_{s_\nu}(t)=E1_{(Z^\mu_t = m_\nu)}(\sigma(t)+q^\zeta(t,U^\zeta_t))$. It remains
to show that the density can be written on the form (\ref{eq:SingleStoppingTimeDensity}).
In order to obtain this, let $Q_{t,\nu}$ be the measure defined in (\ref{eq:gnuetaxMeasure}),
and recall that by Lemma \ref{lemma:gnuetaxSource}, $Q_{t,\nu}$ has a density
given by (\ref{eq:gnuetatDensity}). Note that for all $A\in\BBB$, it holds that
\begin{align}
\int_{E^\nu\times \RR} 1_{(z=m_\nu)}1_A(y)\dv (Z^\mu_t,U^\zeta_t)(P)(z,y)
  &=\int_\RR 1_A(y)\dv Q_{t,\nu}(y),
\end{align}
with $(Z^\mu_t,U^\zeta_t)(P)$ denoting the distribution of $(Z^\mu_t,U^\zeta_t)$. As a consequence,
the same relationship holds with $1_A(u)$ exchanged with arbitrary measurable and bounded
$f:\RR\to\RR$. Applying this, we obtain
\begin{align}
h_{s_\nu}(t) &= E1_{(Z^\mu_t = m_\nu)}(\sigma(t)+q^\zeta(t,U^\zeta_t))\notag\\
&= \int_{E^\mu\times\RR} 1_{(z = m_\nu)}(\sigma(t)+q^\zeta(t,y))\dv (Z^\mu_t,U^\zeta_t)(P)(z,y)\notag\\
&= \int_{\RR} \sigma(t)+q^\zeta(t,y)\dv Q_{t,\nu}(y)
 =\int_0^\infty g_\nu(y|t) (\sigma(t)+q^\zeta(t,y))\dv y,
\end{align}
as was to be shown.
\hfill$\Box$

\textbf{Proof of Lemma \ref{lemma:unuxSource}.} Noting that $(Z^\mu_t = s_\nu)\subseteq(T^\mu_{s_\nu}\le t)$ and
applying Lemma \ref{lemma:SingleStoppingTime}, we obtain
\begin{align}
  u_\nu(t) &= P(Z^\mu_t = s_\nu)
  =P(Z^\mu_t = s_\nu, T^\mu_{s_{\nu}} \le t)\notag\\
  &=\int_0^t P(Z^\mu_t = s_\nu | T^\mu_{s_{\nu}} = v)
             \int_0^\infty g(y|v)(\sigma(v)+ q^\zeta(v,y))\dv y\dv v\notag\\
  &= \int_0^\infty \int_0^t g(y|v) (\sigma(v)+ q^\zeta(v,y)) P(Z^\mu_t = s_\nu | T^\mu_{s_{\nu}} = v) \dv v\dv y\notag\\
  &= \int_0^\infty \int_0^t g(y|v) (\sigma(v)+ q^\zeta(v,y)) \frac{\ell^\gamma_t}{\ell^\gamma_v} \dv v\dv y,
\end{align}
as required.
\hfill$\Box$

\textbf{Proof of Theorem \ref{theorem:gxExpression}.}
Applying the notation and results of Lemma \ref{lemma:gnuetaxSource}, we obtain
\begin{align}
g(t)&=P(Z^\mu_t \in m)
  =\sum_{\nu=1}^\infty Q_{t,\nu}(\RR)
  =\sum_{\nu=1}^\infty \int_0^\infty g_\nu(y|t)\dv y,
\end{align}
as required.
\hfill$\Box$

\textbf{Proof of Theorem \ref{theorem:fetaxExpression}.}
Recalling Lemma \ref{lemma:gnuetaxSource}, it holds for $A\in\BBB_+$ that
\begin{align}
  Q_t(A)
  &=\frac{P(U^\zeta_t \in A,Z^\mu_t = m)}{P(Z^\mu_t = m)}
  =\frac{1}{g(t)}\sum_{\nu=1}^\infty P(U^\zeta_t \in A, Z^\mu_t = m)\notag\\
  &=\frac{1}{g(t)}\sum_{\nu=1}^\infty \int_A g_\nu(y|t)\dv y
  =\int_A \frac{1}{g(t)}\sum_{\nu=1}^\infty  g_\nu(y|t)\dv y,
\end{align}
implying the result.
\hfill$\Box$

\bibliographystyle{amsplain}

\bibliography{full}

\end{document}